\documentclass[twocolumn,pra,aps]{revtex4}

\usepackage{graphicx}
\usepackage{bm}
\usepackage{amsmath}
\usepackage{amssymb}

\def\duzomniejsze{<\kern-.7mm<}
\def\duzowieksze{>\kern-.7mm>}

\def\textbf#1{{\bf #1}}
\def\beq{\begin{equation}}
\def\eeq{\end{equation}}
\def\be{\begin{equation}}
\def\ee{\end{equation}}
\def\ben{\begin{eqnarray}}
\def\een{\end{eqnarray}}
\def\beqa{\begin{eqnarray}}
\def\eeqa{\end{eqnarray}}
\def\eea{\end{array}}
\def\bea{\begin{array}}
\def\bs{{\backslash}}
\newcommand{\bei}{\begin{itemize}}
\newcommand{\eei}{\end{itemize}}
\newcommand{\bee}{\begin{enumerate}}
\newcommand{\eee}{\end{enumerate}}

\def\>{\rangle}
\def\<{\langle}
\def\blacksquare{\vrule height 4pt width 3pt depth2pt}

\def\ot{\otimes}

\newtheorem{lemma}{Lemma}
\newtheorem{corollary}{Corollary}

\newtheorem{theorem}{Theorem}
\newtheorem{definition}{Definition}
\newtheorem{observation}{Observation}

\newtheorem{rem}{Remark}
\newtheorem{example}{Example}

\def\bed{\begin{definition}}
\def\eed{\end{definition}}
\def\bel{\begin{lemma}}
\def\eel{\end{lemma}}
\def\bet{\begin{theorem}}
\def\eet{\end{theorem}}

\begin{document}

\title{On distinguishing of non-signaling boxes via completely locality preserving operations}

\begin{abstract}
We consider discriminating between bipartite boxes with 2 binary inputs and 2 binary outputs ($2\times 2$) using the class of completely locality preserving operations i.e. those, which transform  boxes with local hidden variable model (LHVM) into boxes with LHVM, and have this property even when tensored with identity operation. Following approach developed in entanglement theory we derive linear program which gives an upper bound on the probability of success of discrimination between different isotropic boxes. In particular we provide an upper bound on the probability of success of discrimination between isotropic boxes with the same mixing  parameter. As a counterpart of entanglement monotone we use the non-locality cost. Discrimination is restricted by the fact that non-locality cost does not increase under considered class of operations. We also show that with help of allowed class of operations one can distinguish perfectly any two extremal boxes in $2\times 2$ case and any local extremal box from any other extremal box in case of two inputs and two outputs of arbitrary cardinalities. 
\end{abstract}

\author{K. Horodecki$^{1,2}$}
%04.70.D, 04.40, 05.70
%\homepage[]{Your web page}
%\thanks{}
%\altaffiliation{}
\affiliation{$^1$Institute of Informatics, University of Gda\'nsk, 80--952 Gda\'nsk, Poland}
\affiliation{$^2$National Quantum Information Centre in Gda\'nsk, 81--824 Sopot, Poland}

\maketitle
\section{Introduction}
Asking two distant parties that do not communicate for certain set of answers is the well known scenario for a non-local game \cite{Bell,Brassard-pseudo}. Depending on the resource the parties share, they can obtain higher or lower probability of success in winning the game. Sharing a quantum state can allow for the  probability higher with respect to classical resources, and sharing arbitrary non-local but not-signaling system can make it sometimes even equal to 1 \cite{PR}. 

For this reason, among others, the non-locality represented by the so called box (a conditional probability distribution) has been treated as a resource in recent years \cite{Barret-Roberts}. The world of non-signaling, non-local boxes bares analogy with the world of entangled states \cite{Masanes-NStheor,Pawlowski-Brukner,Ekert91,BHK_Bell_key,Bell-security,Hanggi-phd,BrunnerSkrzypczyk,Allcocketal2009,Forster,Brunneretal2011}.
Therefore it is clear, that investigation of non-locality and entanglement can help each other. We develop this analogy, in the scenario of distinguishing between
systems (see in this context \cite{Bae-dist}). Namely we consider scenario in which two distant parties know that they share a box drawn from some ensemble. Their task is to tell the box they share with as high probability as it is possible, using allowed class of operations. In our case, these operations will be such that transform local boxes into local ones and has this property even when tensored with identity operation. 

An analogous scenario in entanglement theory was considered in recent years (see e.g. \cite{Bandyopadhyay-dist} and references therein), where one asks about discriminating orthogonal quantum states by means of Local Operations and Classical Communication (LOCC). In our method, we base on one of the first results on this subject \cite{BellPRL}, where it was shown that one can not distinguish between 4 Bell states $\{
|\psi_i\>\}_{i=1}^4$ by LOCC. The common method in entanglement theory
is that something is not possible or else some monotone would increase under LOCC operation, which is a contradiction. Here this approach was not directly applicable, since the states (and their entanglement) can be completely destroyed in process of distinguishing. The Ghosh et al. (GKRSS) method of \cite{BellPRL} gets around this problem, by considering entanglement of the Bell states classically correlated with themselves:
\be
\rho_{ABCD}= \sum_i {1\over 4} |\psi_i\>\<\psi_i|_{AB}\ot |\psi_i\>\<\psi_i|_{CD}.
\label{eq:4Bells}
\ee
Indeed, if Alice and Bob could distinguish the Bell states on system AB, they could transform the states on CD into the singlet state by local control operations, that transform each of $|\psi_i\>$ into $|\psi_0\>$ which would mean that distillable entanglement of $\rho_{ABCD}$ is at least 1 e-bit. This contradicts the fact that the state $\rho_{ABCD}$ is separable as it can be written as $\rho_{ABCD}= \sum_i {1\over 4} |\psi_i\>\<\psi_i|_{AC}\ot |\psi_i\>\<\psi_i|_{BD}$. Hence the 
states $\{|\psi_i\>\}$ can not be perfectly distinguishable. 

In what follows, we consider an analogue of the above state (\ref{eq:4Bells}),  based on the so called isotropic boxes $B^{\alpha_i}$ ($B^{\beta_i}$), i.e. boxes that are mixtures of Popescu-Rohrlich (PR) boxes and 'anti' PR box with probability $\alpha_i$ ($\beta_i$). 
\be
B_{in} = \sum_{i=0}^{n-1} p_i B_{AB}^{\alpha_i}\ot B_{CD}^{\beta_i}.
\ee
A class of operations we consider is similar to that of locality preserving \cite{Joshi-broadcasting} but we demand also completeness i.e. that the operations should be locality preserving even when tensored with identity operation on some part of a box. Moreover we demand that they have special output i.e. that they are {\it discriminating} operations. As a monotone we choose a {\it nonlocal cost} of a box \cite{Elitzur-nonloc}. We also show that with help of these operations one can distinguish any 2 extremal boxes in $2\times 2$ case
and any local extremal bipartite box with 2 inputs and 2 outputs from any other extremal box of this form for arbitrary cardinalities of inputs and outputs. This  partially resembles result of \cite{Walgate-twoent} from entanglement theory where it is proven that any two orthogonal (multipartite) states can be perfectly distinguished.

The rest of the paper is organized as follows: section \ref{sec:scenario} provides the scenario and definition of the class of operations. Section \ref{sec:iso-twir-cost} provides useful definitions and some properties of nonlocal cost. In section \ref{sec:bound} we give the main reasoning behind the bound on probability of success in discrimination of isotropic boxes, as well as some corollaries. The proof goes thanks to main inequality:
\be
C(B_{in}) \geq C(B_{out})
\ee 
with $B_{out}$ being $B_{in}$ after discrimination on system $AB$. Finally in section \ref{sec:distinguish} we consider perfect distinguishability of two extremal boxes in bipartite case $2\times 2$ as well as in bipartite case of 2 inputs of arbitrary cardinality and 2 outputs of arbitrary cardinality.

\section{Scenario of distinguishing}
\label{sec:scenario}
By a bipartite box $X$ we mean a family of probability distributions that have support on Cartesian product of spaces $\Omega_{A}\times \Omega_{B}$.
Each of the spaces may contain (the same number of) $n$ systems. 
 In special case of bipartite boxes with $n=1$, we denote them as $P_X(a,b|x,y)$
where $x,y$ denotes the inputs to the box and $a,b$ its output. We say that two boxes are {\it compatible} if they are defined for the same number of parties, with the same cardinalities of each corresponding input and each corresponding output. Definition of multipartite box is analogous.
We will consider only boxes that satisfy some {\it non-signaling} conditions. To specify this we need to define general non-signaling condition between some partitions of systems \cite{Barrett-GPT,Barret-Roberts}.

{\definition Consider a box of some number of systems $n+m$ and its partition into two sets: $A_1,...,A_n$ and $B_1,...,B_{m}$. A box on these systems
given by probability table $P(\bar{a},\bar{b}|\bar{x},\bar{y})$ is non-signaling in cut $A_1,...,A_n$ and $B_1,...,B_{m}$
if the following two conditions are satisfied:
\ben
\forall_{\bar{a},\bar{x},\bar{y},\bar{y}'} \sum_{\bar{b}}P(\bar{a},\bar{b}|\bar{y},\bar{y}) = \sum_{\bar{b}}P(\bar{a},\bar{b}|\bar{y},\bar{y}')\\
\forall_{\bar{b},\bar{x},{\bar{x}'},\bar{y}} \sum_{\bar{a}}P(\bar{a},\bar{b}|\bar{y},\bar{y}) = \sum_{\bar{a}}P(\bar{a},{\bar{b}}|\bar{x}',\bar{y})
\een
If the first condition is satisfied, we denote it as 
\be A_1,...,A_n\not\hspace{-1.3mm}{\rightarrow} B_1,...,B_{m},\nonumber
\ee if the second we write
\be
B_1,...,B_{m}\not\hspace{-1.3mm}{\rightarrow} A_1,...,A_n,\nonumber
\ee and if both:
\be
A_1,...,A_n\not\hspace{-1.3mm}{\leftrightarrow} B_1,...,B_{m}\nonumber .
\ee
We say that A box of systems $A_1,...,A_n,B_1,...,B_{m}$ is fully non-signaling if for any subset of systems $A^IB^J \equiv A_{i_1},...,A_{i_k}B_{j_1},...,B_{j_l}$ with $I\equiv\{i_1,...,i_k\}\subseteq N \equiv\{1,...,n\}$ and $J\equiv\{j_1,....,j_l\}\subseteq M \equiv\{1,...,m\}$ such that not both I and J are empty, there is 
\be A^IB^J\not\hspace{-1.3mm}{\leftrightarrow}A^{N-I}B^{M-J}.
\ee
}

In what follows we will consider only boxes that are fully non-signaling, according to the above definition. The set of all boxes compatible to each other  
, that satisfy the above definition, we denote as $NS$.

By locally realistic box we mean the following ones:
{\definition Locally realistic box of $2n$ systems $A_1,...,A_n,B_1,...,B_n$ is defined as 
\be
\sum_\lambda p(\lambda)P{(\bar{a}|\bar{x})}_{A_1,...,A_n}^{(\lambda)}\otimes P{(\bar {b}|\bar {y})}_{B_1,...,B_n}^{(\lambda)}
\label{eq:local}
\ee
for some probability distribution $p(\lambda)$, where we assume that 
boxes $P{(\bar{a}|\bar{x})}_{A_1,...,A_n}^{(\lambda)}$ and 
$P{(\bar{b}|\bar{y})}_{B_1,...,B_n}^{(\lambda)}$ are fully non-signaling. 
The set of all such boxes we denote as $LR_{ns}$. All boxes that are fully non-signaling
but do not satisfy the condition (\ref{eq:local}), are called non-$LR_{ns}$.
}

We consider a family $\cal L$ of operations $\Lambda$ on a box shared between Alice and Bob, which preserve locality, as defined below (see in this context \cite{Joshi-broadcasting}).

{\definition An operation $\Lambda$ is called locality preserving (LP) if it satisfies the following conditions:

(i) {\it validity} i.e. transforms boxes into boxes.

(ii) {\it linearity} i.e. for each mixture $X = p P + (1-p) Q$, there is $\Lambda(X) = p \Lambda(P) + (1-p) \Lambda(Q)$

(iii) {\it locality preserving } that is transforms boxes from $LR_{ns}$ into boxes from $LR_{ns}$.

(iv) {\it non-signaling} that is transforms fully non-signaling boxes into fully non-signaling ones.
\label{def:LP} 
}

In what follows we will focus on special locality preserving operations, namely those which are completely locality preserving.

{\definition An operation $\Lambda$ acting on system $AB$ is called completely locality preserving (CLP) if $\Lambda$ is locality preserving and 
$\Lambda \ot I_{CD}$ is locality preserving where $I_{CD}$ is identity operation on arbitrary but finite-dimensional bipartite system of subsystems $C$ and $D$.
}

{\rem Note, that $CLP \subsetneq LP$, since the swap operation $V$ is in LP, but is not in CLP: $V_{AB}\otimes I_{A'B'}$ acting on product of two nonlocal boxes on $AA'$ and $BB'$ respectively, makes non-locality across $AA'$ versus $BB'$ cut. Similarly like the swap operation on quantum states transforms separable states into separable ones, yet is not completely separable to separable state operation, as $V_{AB}\otimes I_{A'B'}$ creates entanglement in $AA'$ vs $BB'$ cut, when applied to tensor product of two singlet states: $|\psi_0\>_{AA'}\otimes|\psi_0\>_{BB'}$.
}

Finally, we will be interested in those CLP maps which are discriminating between some boxes from a given {\it ensemble}, where by ensemble we mean the
family of pairs $\{p_i,X_i\}_{i=0}^{n-1}$ where $X_i$ is a bipartite box, and $p_i$ is the probability with which Alice and Bob share this box such that $\sum_{i=0}^{n-1} p_i = 1$. We will need
also the notion of a {\it flag-box} which is a box denoted as $F(j)$ defined as deterministic box with single input $s$ of cardinality 1, 
and as a (single) output a probability distribution on $\{1,...,n\}$ which is  Kronecker delta $\delta_{j,e}$. To indicate its input and output, we will denote it also as $P^{(j)}(e|s)$. It can be viewed as a counterpart of quantum state $|j\>\<j|$, and it is equivalent to probability distribution \cite{Short-Wehner}. We say, that an $F(j)$ is a flag-box with flag $j$. In what follows an operation returning flag-boxes with flag $j$ means that Alice and Bob claim that they were given box number $j$ from the ensemble.

{\definition $\Lambda$ discriminates the ensemble $\{p_i,X_i\}_{i=0}^{n-1}$ if for every $i$, there is
\be
\quad \Lambda(X_i) = \sum_{j=0}^{n-1}q^{X_i}_{j}F^A(j)\ot F^B(j)
\ee
where $\{q^{X_i}_{j}\}_{j=1}^n$ is a probability distribution that may depend on $X_i$. The box $F^A(j)$ is a flag-box with flag $j$ on system of Alice and $F^B(j)$ is that on Bob's.
\label{def:discrim}
}

Note, that the above definition could be defined without reference to ensemble: just on any box $X$ discriminating operation should provide flag-boxes. However we find the latter, in principle more restrictive one. 

\begin{figure}[!h]
\includegraphics[width=5cm]{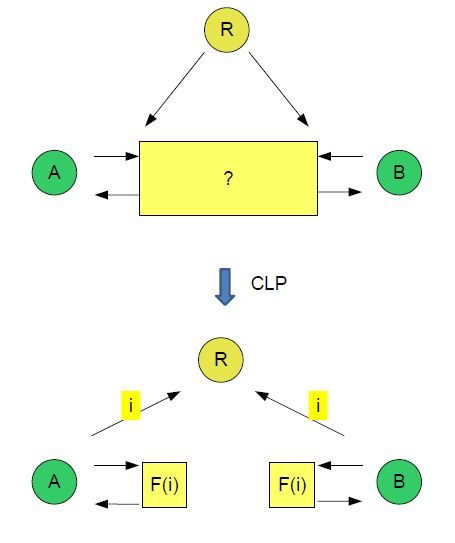}
\caption{ Depiction of the considered scenario: Alice and Bob are give by the Referee R one of the boxes $B^{\alpha_i}_i$ with probability $p_i$. They apply CLP operation to distinguish between them, and send the guess $i$ to the Referee}
\label{fig:scenario}
\end{figure}

We can describe now the scenario of discrimination of an ensemble. The Referee creates a box on systems $RAB$ of the form:
\be
\sum_{i=0}^{n-1} p_i F^R(i)\ot X^{AB}_i
\ee
and then sends system $A$ to Alice and $B$ to Bob, distributing thereby between them the box $X_i$ with probability $p_i$. The Referee holds flag-box $F(i)$ and
waits for their answer. Alice and Bob are allowed to apply some operation which is (i) CLP and (ii) discriminates the ensemble $\{p_i,X_i\}_{i=0}^{n-1}$, denoted as $\Lambda$. Due to Definition \ref{def:discrim}, by linearity of CLP operations, $\Lambda$ results in the following box shared between the Referee, Alice and Bob (see Fig. \ref{fig:scenario}):

\be
\sum_{i=0}^{n-1}\sum_{j=0}^{n-1}p_i q^{(i)}_{j}F^R(i)\ot F^A(j)\ot F^B(j)
\ee

We define now the {\it probability of success $p_{\Lambda}^s$ with which $\Lambda$ discriminates the ensemble}. 
It is computed from the joint probability distribution of the Referee's 'flags' and the Alice's and Bob's 'flags' 
$p_{\Lambda}(i,j)\equiv p_i q^{(i)}_{j}$ as
\be
p_{\Lambda}^s \equiv \sum_{i=0}^{n-1} p_{\Lambda}(i,i)
\ee

We can finally define the problem of distinguishing between boxes as follows: 

{\it Given an ensemble of bipartite non-signaling boxes
\be
\{p_i, X_i\},
\ee
find the maximal value of probability of success $p_s$ in discriminating between the given boxes using operations CLP that discriminates the ensemble}. 

One can be interested if the set of $CLP$ operations that distinguishes an ensemble is not empty. It is easy to observe, that any composition of local operations on both sides is a valid CLP operation providing the local operations satisfy non-signaling condition. It is however not easy to see, if such operations could produce the same flag-boxes for Alice and for Bob, which are correlated with given ensemble \footnote{ To obtain flag-boxes uncorrelated with the ensemble is easy if we allow share randomness, but even with this resource it is not clear if demanding the output of the form of the same flag-boxes is not too rigorous}. 
However, as we show in the Appendix \ref{subsec:comparing-ops}, the following operation is CLP operation which discriminates the ensemble: it is a composition of  
(i) local measurements, (ii) exchanging the results, (iii) grouping them into disjoined sets  and (iv) creating the same flag-boxes for each group followed by tracing out of the results of measurements (see example below). We will call this type of operation the {\it comparing operations} as the parties decide on the guess after comparing their outputs. We note here, that output of the form of the same two  flag-boxes is crucial for further considerations, as thanks to having the same flag-boxes, both Alice and Bob can transform their box conditionally on output of distinguishing.

{\example Consider a pair of boxes: the PR box, defined as

\be P_1(a,b|x,y) = \left\{ \begin{array}{ll}
         {1\over 2} & \mbox{if $a\oplus b = xy$} \\
        0 & \mbox{else},\end{array} \right.
\ee
and anti-PR box defined as
\be P_2(a,b|x,y) = \left\{ \begin{array}{ll}
         {1\over 2} & \mbox{if $a\oplus b = xy\oplus 1$} \\
        0 & \mbox{else}.\end{array} \right.
\ee
Then, by (i) choosing $x=1$ (Alice) and $y=1$ (Bob), comparing the results (ii) deciding to output flags $F(1)^A\otimes F(1)^B$ if the results are not equal (and hence
$a\oplus b = 1$ while  $F^A(2)\otimes F^B(2)$ providing the results are equal (and hence $a\oplus b = 0$) (iii) tracing out the results of measurements, they distinguish perfectly the PR box from anti-PR box via a CLP operation.
}

\section{Isotropic boxes, twirling and non-local cost}
\label{sec:iso-twir-cost}
In what follows, we will use numerously the boxes locally equivalent to PR box:

\be B_{rst}(a,b|x,y) = \left\{ \begin{array}{ll}
         1/2 & \mbox{if $a\oplus b = xy\oplus rx \oplus sy \oplus t$} \\
        0 & \mbox{else}.\end{array} \right.
\ee
(where $a,b,x,y,r,s,t$ are binary), which we call here maximally nonlocal boxes.

\begin{figure}[!h]
\includegraphics[width=7cm]{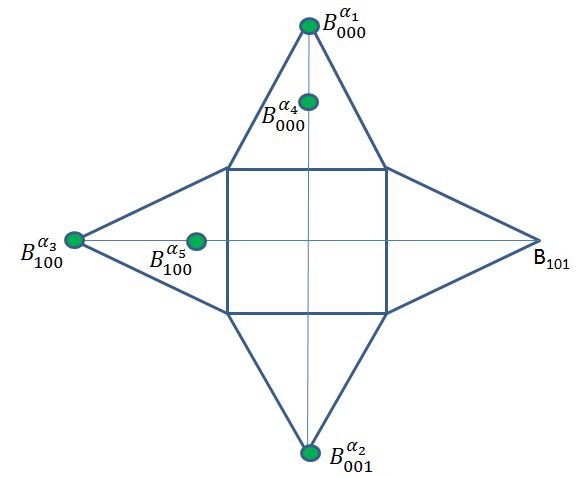}
\caption{Exemplary ensemble $\{{1\over 5},B^{\alpha_i}_{i}\}_{i=1}^5$. The members of ensemble are depicted as green circles. The square
depicts the set of local boxes. The function $f$ is defined as: $f(1)=000, f(2)=001, f(3)=100, f(4)=000, f(5)=100$ and there is $\alpha_1=\alpha_2=\alpha_3=1$, $\alpha_4 = {1\over 6}$ and $\alpha_5 = {1\over 4}$. }
\label{fig:fig1}
\end{figure}

More specifically, we will focus on distinguishing between {\it isotropic} boxes \cite{Short} 
\be
B_i^{\alpha_i} = \alpha_i B_i + (1-\alpha_i) \bar{B}_i.
\ee
$B_i\in\{B_{rst}\}_{rst=000}^{111}$, where $\alpha_i \in (3/4, 1]$,
and $\bar{B}_i$ denotes $B_{rs\bar{t}}$ with $\bar t$ being negation of bit t.
We define here a function $f$ which maps indices $i$ into strings $rst$, that is
$f(i):=rst$ iff $B_i^{\alpha_i} = \alpha_i B_{rst} + (1-\alpha_i)B_{rs\bar{t}}$. In other words, this 
function groups isotropic boxes according to which maximally nonlocal box it is built of. 
By $B_{rst}^{\alpha_i}$ we will denote $B_i^{\alpha_i}$ such that 
$f(i) = rst$ i.e. $\alpha_i B_{rst} + (1-\alpha_i)B_{rs\bar{t}}$. We exemplify this notation on Fig. \ref{fig:fig1}

The boxes $B_{000}$ and $B_{001}$ are invariant under the following transformation \cite{Short,Nonsig-theories} called twirling:
{\definition A twirling operation $\tau$ is defined by flipping randomly 3 bits $\delta,\gamma,\theta$ and applying the following transformation to a 2x2 box:
\ben
x &\rightarrow & x \oplus \delta \nonumber \\
y &\rightarrow &y \oplus \gamma \nonumber \\
a &\rightarrow &a \oplus \gamma x \oplus \delta\gamma \oplus \theta  \nonumber \\ 
b &\rightarrow &b \oplus \delta y \oplus \theta  \nonumber \\
\een
\label{def:twirling}
}

In what follows, as a measure of non-locality we take the non-locality cost $C(P)$ \cite{PR,Brunneretal2011} defined like this:
\be
C(P) = \inf \{p: P= pX + (1-p)L,\, X \in NS,\, L \in LR_{ns}\}.
\ee
We make the following easy observation, that non-locality cost is monotonous under CLP operations.
{\observation
Nonlocality cost does not increase under LP and CLP operations
\label{obs:monot}
}

{\it Proof}.-
Let us consider any ensemble of $P = p X + (1-p)L$. By linearity of $\Lambda$
\be
\Lambda(P) = p\Lambda(X) + (1-p)\Lambda(L).
\ee
Now, no matter $\Lambda$ is LP, or CLP, it is locality preserving which implies that $\Lambda(L)$ is some $L'$ from $LR_{ns}$. 
Moreover it is valid, hence transforms $X$ into some non-signaling box $\Lambda(X)$.
\be
\Lambda(P)= p\Lambda(X) + (1-p)L'.
\ee
Hence, there is $p \geq C(\Lambda(P))$ since this valid decomposition into local $L'$ 
and nonlocal part $\Lambda(X)$ can be suboptimal for $C(\Lambda(P))$. Since this happens for any ensemble, and $C(P)$ 
is infimum of $p$ over the ensembles, we have that $C(\Lambda(P))\leq C(P)$, by definition of infimum. Indeed,
for any $\delta$ there exists $p_{\delta}$ which is such that $C(P) + \delta > p_{\delta}$ thus by contradiction
if $C(\Lambda(P))> C(P)$ then taking $\delta = C(P)-C(\Lambda(P))$ we would get $C(\Lambda(P)) > p_{\delta}$, which
contradicts the above considerations.
\blacksquare

We will need also an observation (see in this context \cite{Brunneretal2011,Joshi-broadcasting})

{\observation An isotropic box $B^{\alpha}_{000}= \alpha B_{000} + (1-\alpha) B_{001}$ with $\alpha \in ({3\over 4},1]$ satisfies
\be
C(B^{\alpha}_{000})=4\alpha -3. 
\ee
\label{obs:iso}
}

{\it Proof}.- Since the box $B^{\alpha}_{000}$ is invariant under twirling, it's optimal decomposition in definition of $C(P)$
has both local part $L$ and nonlocal $X$ which are also invariant under twirling i.e. lays on a line between $B_{000}$ and $B_{001}$. 
Let us consider some decomposion $B^{\alpha}_{000} = p X + (1-p) L$, where $L$ is a local box.  
Note, that $p$ in this decomposition can be written in terms of CHSH value:
\be
\gamma(X) = \<00\> + \<01\> + \<10\> -\<11\>,
\ee
with $\<ij\>=P(a=b|x=i,y=j)-P(a\neq b|x=i,y=j)$. Namely:
\be
p = {\gamma(B^{\alpha}_{000}) -\gamma(L) \over \gamma(X) - \gamma(L)}.
\ee
It is now easy to see, that for fixed $L$, minimal $p$ is reached for $X = B_{000}$, as we can always lower the $p$ by setting 
$\gamma(X) = 4$. Hence we end up with optimization of a function
\be
{(8\alpha -4) -\gamma(L) \over 4 - \gamma(L)},
\ee
where $-2 \leq \gamma(L)\leq 2$. Using Mathematica 7.0, we find this function attains minimum
at $4\alpha -3$, which we aimed to prove.\blacksquare

\section{Upper bound on distinguishing of isotropic boxes}
\label{sec:bound}
\begin{figure}[!h]
\includegraphics[width=7cm]{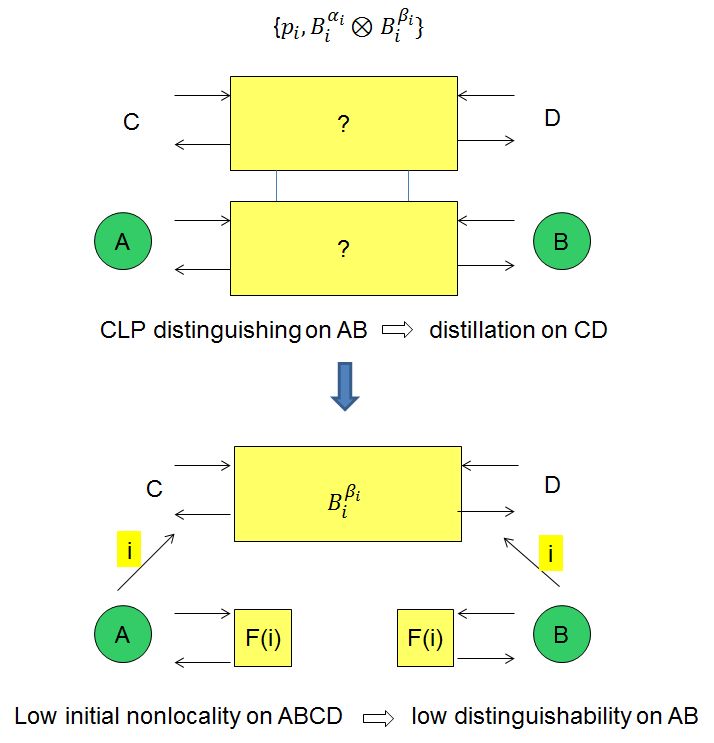}
\caption{Illustration of the analogue of the GKRSS method: Alice and Bob could apply the CLP distinguishing operation to the box $B_{in}=\sum_{i=0}^{n-1}p_i B_i^{\alpha_i}\ot B_i^{\beta_i}$ and via distinguishing
on AB distill boxes $B^{\beta_i}_i$ on CD. If the initial non-locality of $B_i$ is small, the success in distinguishing is limited, as distillation can not
exceed initial cost of non-locality of $B_i^{\alpha_i}$.}
\label{fig:fig-GKRSS}
\end{figure}

We focus now on distinguishing of the following ensemble:
\be
\{p_i, B_i^{\alpha_i}\}_{i=0}^{n-1}.
\ee
with $\alpha_i \in [{1\over 2},1]$.
Following GKRSS method \cite{BellPRL}, we will consider a box obtained by classically correlating boxes $B_i^{\alpha_i}$ with other isotropic boxes, parametrised by some $\beta_i \in [{1\over 2},1]$.
\be
B_{in}=\sum_{i=0}^{n-1}p_i B_i^{\alpha_i}\ot B_i^{\beta_i},
\ee
and compare its non-locality with the box after application of some optimal CLP discriminating operation $\Lambda$ (see Fig. \ref{fig:fig-GKRSS}). 

We obtain the following result:

{\theorem For an ensemble $\{p_i,B_i^{\alpha_i}\}_{i=0}^{n-1}$ with $\alpha_i \in [{1\over 2},1]$ and any CLP operation $\Lambda$ which discriminates the ensemble 
there is
\be
\sum_{i=0}^{n-1} p_{\Lambda}(i,i)(\beta_i +\max_k \beta_k -1) \leq {C(B_{in}) + 3\over 4} + \max_k\beta_k-1,
\ee
where $B_{in} = \sum_{i=0}^{n-1} p_i B_i^{\alpha_i}\ot B_i^{\beta_i}$.
\label{thm:main}
}

Following the above theorem we have immediate corollary, considering distinguishing of isotropic states with the same parameter $\alpha_i=\alpha$, and considering
$\beta_i = \alpha$.

{\corollary For the ensemble of isotropic boxes with the same parameter $\alpha_i =\alpha\in [{3\over 4},1]$: $\{p_i,B_i^{\alpha}\}$ with $n\leq 8$ 
the optimal probability of distinguishing by CLP operations that discriminate the ensemble satisfies:
\be
p_s \leq {C(B_{in}) -1 + 4\alpha\over 4(2\alpha -1)},
\ee
where $B_{in} = \sum_{i=0}^{n-1}  p_i B_i^{\alpha}\ot B_i^{\alpha}$.
\label{cor:alpha}
}

The main corrolary concerns discriminating between the boxes while setting $\beta_i =1$ for all $i$ i.e. setting $B^{\beta_i}_i$ to be maximally non-local:

{\corollary For the ensemble of maximally nonlocal boxes $\{p_i,B^{\alpha}_i\}$ with $n\leq 8$, the optimal probability of distinguishing by CLP operations that discriminate the 
ensemble satisfies:
\be
p_s \leq {C(B_{in}) + 3 \over 4},
\ee
where $B_{in} = \sum_{i=0}^{n-1} p_i B^{\alpha}_i\ot B^{1}_i$.
\label{cor:nonlock-flags}
}

To exemplify the consequences of the above corollary we first set also $\alpha =1$, that is consider distinguishing between maximally non-local boxes.
We consider then the ensembles with a fixed number of maximally non-local boxes $k$ provided with equal weights $p_i={1\over k}$, and for each of them find $C(B_{in})$. We then take minimal of these values, obtaining universal bound $p_s(k)$ - on the probability of success of distinguishing $k$ maximally non-local boxes from each other. 

To this end, we have used Mathematica 7.0 and approach of \cite{Brunneretal2011}, but with much smaller class of deterministic boxes, since we demand
stronger non-signaling condition. For $k=2$ the bound is trivial, as the cost is 1 for any pair. For $k=3$ there are only 6 ensembles for which cost is less 
than one (equal $1\over 3$ for all 6) which implies the bound ${11\over 12}$ for all of them. For example 
\be
A_3=\{{1\over 3}B_{000},{1\over 3}B_{010},{1\over 3}B_{100}\},
\ee
has $p_s(A_3)\leq {11\over 12}$. 
For $k=4$ an exemplary ensemble with the smallest non-locality cost $5\over 8$ is
\be
A_4 = \{{1\over 4}B_{000},{1\over 4}B_{001},{1\over 4}B_{010},{1\over 4}B_{100}\}
\ee
and there is $p_s(A_4)\leq {29\over 32}$. 
For $k\geq 5$ the non-locality cost is non-zero for every box that we consider and hence we have obtained the following general bounds: 

\ben
p_s(5) \leq {37\over 40}, \nonumber \\
p_s(6) \leq {7\over 8},  \nonumber \\
p_s(7) \leq {23\over 28}, \nonumber \\
p_s(8) \leq {3\over 4}. \nonumber 
\een

One can be also interested if some bound can be obtained for a box which can be obtained physically, i.e. via measurement on a quantum state. We choose the boxes
with $\alpha_i = {2 + \sqrt{2} \over 4}=\alpha_q$, which corresponds to CHSH quantity equal to $2\sqrt{2}$ \cite{CHSH,Tsirelson}. We obtain via corollary \ref{cor:nonlock-flags}, 
denoting $p_s^{\alpha_q}(k)$ the upper bound on probability of success of discriminating between any $k$ boxes from the set $B^{\alpha}_{rst}$ that
\ben
p_s^{\alpha_q}(3) \leq 0.975593, \nonumber \\
p_s^{\alpha_q}(4) \leq 0.926778, \nonumber \\
p_s^{\alpha_q}(5) \leq 0.874817, \nonumber \\
p_s^{\alpha_q}(6) \leq 0.833334, \nonumber \\
p_s^{\alpha_q}(7) \leq 0.785715, \nonumber \\
p_s^{\alpha_q}(8) \leq 0.750001, \\
\een
where we rounded numerical results at $6$th place.
Interestingly, although corollaries \ref{cor:alpha} and \ref{cor:nonlock-flags} are not directly comparable as they have different $B_{in}$, in this case corollary
\ref{cor:alpha} leads to worse result than above, in particular $p_s^{(3)}$ in that case is bounded by more than 1. 

\subsection{ Proof of theorem \ref{thm:main}}

Before we prove theorem \ref{thm:main}, we need to make some necessary observations.
We will compare the initial value of non-local cost of a box with its value after applying distinguishing operation
and special post-processing. The box $B_{in}$ after distinguishing equals
\be
B_{out}=\sum_{i,j=0}^{n-1} p_{\Lambda}(i,j) F^A(j)\otimes F^B(j)\otimes B_i^{\beta_i}.
\label{eq:out-box}
\ee
To this box we apply a post-processing transformation which is composition of (i) local reversible control-$O_j$ operation that is operation of certain rotation $O_j$ of $B_i^{\beta_i}$ controlled by index $j$ of $F(j)^A$ followed by (ii) application of the twirling $\tau$ of the target system (iii) tracing out of the control system.
The role of the (i) operation is to use the fact that if Alice and Bob would discriminate well the boxes $B_i^{\alpha_i}$, then they would obtain on other system
by control operation a box that has high non-locality cost (the $O_j$ rotations are such that resulting state has large fraction of PR box, like in GKRSS method, a singlet was obtained). The operations (ii) and (iii) has only technical meaning: they map the resulting box $B_{out}$ into $2\times 2$ isotropic box $B'_{out}$, so that we are able to calculate the non-locality cost for this box via observation \ref{obs:iso}, and hence lower bound non-locality cost of $B_{out}$.

After applying operations (i)-(iii), the output box is of the form
\be
B'_{out}=\tau(\sum_{i,j=0}^{n-1} p_{\Lambda}(i,j)O_j(B_i^{\beta_i})).
\ee
where $O_j(B_i^{\beta_i})$ is an operation defined such that for $f(j)= rst$ and $f(i)=r's't'$ 
\be
O_j(B^{\beta_i}_i) \equiv B^{\beta_i}_{(r\oplus r'),(s\oplus s'),(r's\oplus s'r\oplus t\oplus t')}, 
\ee
and it is a local operation: some combination of flipping (or not) inputs x,y
and output b \footnote{ More specifically, $O_j(B_{rst}^{\alpha_i})$ for $f(j)=000$ acts as identity, for $f(j)=001$ it is a b-flip (negation of output),
for $f(j)=010$ is x-flip (negation of input x), for $f(j)=011$ is x-flip and b-flip, for $f(j)=100$ is y-flip (negation of input y) and for $f(j)=101$ is y-flip with b-flip.
Finally for $f(j)=110$ it is both x-flip and y-flip, while for $f(j)=111$ it is both x and y-flip with b-flip.}.

We make now some necessary observations.

{\observation For a valid, linear operation $\Lambda$ which maps a non-signaling box $B$ into non-signaling box $\Lambda(B)$,
and transforms $LR_{ns}$ boxes into $LR_{ns}$ there is:
\be
C(B) \geq C(\Lambda(B))
\ee
\label{obs:particular}
}

{\it Proof}. The proof of this fact goes in full analogy to proof of observation \ref{obs:monot}. The only difference is that
the operation $\Lambda$ may not possess all properties of CLP operation for other boxes than $B$ and $\Lambda(B)$.\blacksquare

{\corollary The composition of operations of control-$O_j$, twirling on target system
and tracing out control system applied to box (\ref{eq:out-box}) does not increase the non-locality cost.
\label{cor:loc-pres}
}

{\it Proof}.  It is easy to check (see Appendix \ref{app:proof} for full argument) that a box of the form
(\ref{eq:out-box}) and control-$O_j$ operation satisfy assumptions of the observation \ref{obs:particular}, while twirling and tracing out of a subsystem
are just CLP operations, hence the composition of those three can not increase cost of non-locality.\blacksquare

From definition of $O_j$ operation, there follows directly an observation:

{\observation The operations $O_j$ satisfy the following relations:
\be
O_j(B_i^{\beta_i}) = B_{000}^{\beta_i} \quad \mbox{for $f(j)=f(i)$}.
\ee
Moreover for all $0 \leq i,j\leq n-1$ there is $O_j(B_i^{\beta_i})=B_{rst}^{\beta_i}$ for some
$rst\in\{000,...,111\}$. 
\label{obs:rotations}
}

We will need now the following observation concerning twirling operation:

{\observation For any $0\leq i \leq n-1$ there is 
\be
\tau(B_i^{\beta_i}) = B_i^{\beta_i} \quad \mbox{for} \quad f(i)=000,001
\ee
and 
\be
\tau(B_i^{\beta_i})= {1\over 2}(B_{000} + B_{001}) \quad \mbox{for} \quad f(i)\in\{010,...,111\}.
\ee
where $\tau$ is the twirling operation given in def \ref{def:twirling}.
\label{obs:twirling}
}

{\it Proof}. Follows directly from definition of the twirling operation and 
the boxes $B_{rst}$.\blacksquare

We can now pass to prove theorem \ref{thm:main}.

{\it Proof}.
By monotonicity under locality preserving operations (observation \ref{obs:monot}), the fact that $B_{out}$ is a result of 
CLP map, and corollary \ref{cor:loc-pres} we have 
\be
C(B_{in}) \geq C(B_{out}) \geq C(B'_{out}),
\ee
hence, to prove the thesis it suffice to show that there is 
$C(B'_{out}) \geq 4[\sum_i p_{\Lambda}(i,i)(\beta_i +\max_k \beta_k -1) + (1-\max_k \beta_k)] - 3$. Thus by observation \ref{obs:iso}, it suffice
to show that if we decompose $B'_{out}$ as $q B_{000} + (1-q)B_{001}$, the mixing parameter $q$ will satisfy
\be
q \geq \sum_{i=0}^{n-1} p_{\Lambda}(i,i)(\beta_i +\max_k \beta_k -1) + (1-\max_k \beta_k).
\label{eq:aim}
\ee 
Recall that 
\be
B'_{out}=\tau(\sum_{i,j=0}^{n-1} p_{\Lambda}(i,j)O_j(B_i^{\beta_i})).
\ee
This by linearity of twirling and observation \ref{obs:rotations} equals
\be
B'_{out}=\sum_{i=0}^{n-1}p_{\Lambda}(i,i)B_{000}^{\beta_i} + \sum_{i\neq j}^{n-1} p_{\Lambda}(i,j)\tau(B_{j|i}^{\beta_i}).
\ee
with $j|i \in \{000,...,111\}$. 
Now by observation \ref{obs:twirling} there is 
\ben
B'_{out}=\sum_{i=0}^{n-1}p_{\Lambda}(i,i)B_{000}^{\beta_i} + \nonumber \\
\sum_{i\neq j}^{n-1} p_{\Lambda}(i,j)[u_{ij}B_{000} + (1- u_{ij})B_{001}].
\een
where $u_{ij} = 1/2$ for $i,j$ such that $j|i\neq 000$ and $j|i\neq 001$ while
$u_{ij}= \beta_i$ for all i and j such that $j|i = 000$ and $u_{ij}=(1-\beta_i)$
for all i and j such that $j|i = 001$.
Hence the multiplying coefficient of $B_{000}$ reads
\ben
q =\sum_{i=0}^{n-1} p_{\Lambda}(i,i)\beta_i + \sum_{i\neq j, j|i\neq 000, j|i\neq 001} p_{\Lambda}(i,j){1\over 2} + \nonumber \\ 
\sum_{i, j|i = 000}p_{\Lambda}(i,j)\beta_i + \sum_{i, j|i = 001}p_{\Lambda}(i,j)(1-\beta_i).
\een 
Since we have $\beta_i \in [{1\over 2}, 1]$ there is $(1-\beta_i) \leq \beta_i$  and $(1-\beta_i)\leq {1\over 2}$.
Thus there is
\be
q \geq \sum_{i=0}^{n-1} p_{\Lambda}(i,i)\beta_i + \sum_{i\neq j} p_{\Lambda}(i,j)(1-\beta_i).
\ee
and further
\be
q \geq \sum_{i=0}^{n-1} p_{\Lambda}(i,i)\beta_i + \sum_{i\neq j} p_{\Lambda}(i,j)(1-\max_k \beta_k),
\ee
which is nothing but
\be
q\geq \sum_{i=0}^{n-1} p_{\Lambda}(i,i)\beta_i + (1-\sum_i p_{\Lambda}(i,i))(1-\max_k \beta_k).
\ee
which is equivalent to (\ref{eq:aim}), and the assertion follows.\blacksquare

\section{Discriminating between extremal boxes}
\label{sec:distinguish}
In this section, we  apply the {\it comparing operations} to distinguish some boxes perfectly, i.e. with $p_s = 1$. More precisely, we show that in $2\times 2$ case any extremal boxes are distinguishable by comparing operations. We also prove, that in case of 2 inputs and 2 outputs, whatever the cardinality of inputs and outputs, any local extremal box is distinguishable from any extremal box, by these operations.

{\observation For any two bipartite boxes $X_1 \neq X_2$ compatible with each other, there is a lower bound on probability of success in distinguishing them when provided with equal probabilities, via CLP operation:
\be
{1\over 2} + {1\over 4}\max_{x,y}[\sum_{a,b} |P_{X_1}(a,b|x,y) - P_{X_2}(a,b|x,y)|]
\ee
} 

{\it Proof}. The proof is due to the fact that comparing operations given in eq. (\ref{eq:exCLP}) of Appendix, are CLP. This means that the parties can choose the best measurement $(x,y)$ and then group the results according to Helstrom optimal measurement \cite{Hayashi-book}, which attains the variational distance between the conditional probability distributions $P_{X_1}(a,b|x,y)$ and $P_{X_2}(a,b|x,y)$.\blacksquare

We now turn to special case, where we discriminate only between extremal boxes. The intuition is that they should be to some extent distinguishable, and this
is the case as we show below. We first focus on $2\times 2$ case because they are extremal (similarly like pure quantum states). In what follows, by support of a box $E$, we will mean the following set:
\be
\mbox{supp}E \equiv \{(a,b,x,y): P_E(a,b|x,y) > 0\}.
\ee

{\theorem Any two $2\times 2$ extremal boxes are perfectly distinguishable by some CLP operation.
}

{\it Proof}. 

It is easy to see that each local boxes are distinguishable among others since by locality they need 
to have disjoined support of some probability distributions, and measuring this probability distribution 
determines which local box we have. For other cases the proof boils down to checking that there always exists
a pair of entries $x$ and $y$ such that the resulting probability distributions for two extremal boxes have disjoined support.
Hence, upon a {\it comparing operation} which starts from measuring this pair of entries, the boxes are perfectly distinguishable.
\blacksquare

In order to partially generalize this result to the case of larger dimensions, we now observe general property of extremal boxes: support of one
can not be contained in the support of the other or else the latter would not be extremal. 

{\lemma For any two extremal $n$-partite boxes $E_1 \neq E_2$ of the same dimensionality, there is $\mbox{supp} E_1 \nsubseteq \mbox{supp} E_2$.
\label{lem:ext-dis}
 }
  
{\it Proof}. For clarity, we state the proof for a bipartite boxes, since that for $n$-partite, following similar lines. Suppose by contradiction, that $\mbox{supp} E_1 \subseteq \mbox{supp} E_2$. Then if all probabilities of $E_1$ are less than or equal to corresponding probabilities of $E_2$ (for every measurement), then $E_1 = E_2$. Indeed, if there was some $(a_0,b_0,x_0,y_0)$ such that
\be
P_{E_1}(a_0,b_0|x_0,y_0) < P_{E_2}(a_0,b_0|x_0,y_0),
\ee
then 
\be
\sum_{a,b} P_{E_1}(a,b|x_0,y_0) < \sum_{a,b} P_{E_2}(a,b|x_0,y_0) = 1,
\ee
which is a contradiction since $\{P_{E_1}(a,b|x_0,y_0)\}$ is a probability distribution. 
Thus we may safely assume that there exists $(a_0,b_0,x_0,y_0)$
such that 
\be
P_{E_1}(a_0,b_0|x_0,y_0) > P_{E_2}(a_0,b_0|x_0,y_0).
\ee
Let us denote $T =\{P_{E_2}(a,b|x,y): P_{E_2}(a,b|x,y) \leq P_{E_1}(a,b|x,y)\}$, and
$S=\{P_{E_1}(a,b|x,y): P_{E_2}(a,b|x,y) \leq P_{E_1}(a,b|x,y)\}$
By the above consideration we have that
\ben
r_1\equiv\min_{(a,b,x,y)\in \mbox{supp} E_2} T
\een
is well defined, and by definition satisfies $r_1 >0$.
Moreover, for
\ben
r_2\equiv \max_{(a,b,x,y)\in \mbox{supp} E_1} S
\een
there is $r_2 > r_1$ as it follows from: $r_2 \geq P_{E_1}(a_0,b_0|x_0,y_0) > P_{E_2}(a_0,b_0|x_0,y_0)\geq r_1$. 
By positivity of $r_1$ and from the above inequality we have $p\equiv {r_1\over r_2}$ satisfies $0< p < 1$ i.e. it can be interpreted as non-trivial probability. 
This however gives, that
\be
\tilde{E} \equiv {E_2 - pE_1\over {1-p}}
\ee
is a valid box. Indeed, for all $(a,b,x,y)$ there is
\be
P_{E_1}(a,b|x,y) {r_1\over r_2} \leq P_{E_2}(a,b|x,y)
\label{eq:order}
\ee
since either $P_{E_1}(a,b|x,y) \leq P_{E_2}(a,b|x,y)$, and then ${r_1 \over r_2}<1$ gives the above inequality, or
$P_{E_1}(a,b|x,y) > P_{E_2}(a,b|x,y)$ and then $P_{E_1}(a,b|x,y) \in S$ and $P_{E_2}(a,b|x,y) \in T$. In the latter case, by definition of $S$ there is
$P_{E_1}(a,b|x,y){1\over r_2} \leq 1$, while $P_{E_2}(a,b|x,y) \geq r_1$ by definition of $T$, which proves (\ref{eq:order}).

The box $\tilde{E}$ is also non-signaling, as a difference of two (unnormalized) non-singalling boxes. In turn, there is:
\be
E_2 = p E_1 + (1-p) \tilde{E}
\ee
hence $E_2$ is a non-trivial mixture of two non-signaling boxes. This is desired contradiction, since $E_2$ is by assumption an extremal box, hence the assertion follows. \blacksquare

To state the result that follows from the above lemma, we need a definition of conclusive distinguishing:

{\definition We say that a multipartite box $X$ can be conclusively distinguished from a multipartite box $Y$ compatible with $X$, with nonzero probability if for there exists measurement $x^{0}_1,...,x^{0}_n$ such that there exist(s) outcome(s) $(a^{i}_1,...,a^{i}_n)$ for which $p=\sum_i P_X(a_1^{i},...,a_n^{i}|x^{0}_1,...,x^{0}_n) > 0 $  but $P_Y(a_1^{i},...,a_n^{i}|x^{0}_1,...,x^{0}_n) = 0$ for all $i$. We then say that $X$ is conclusively distinguishable from $Y$ with at least probability $p$.
}

From lemma \ref{lem:ext-dis} it direcly follows that

{\theorem For any two extremal multipartite boxes $E_1 \neq E_2$ of the same dimensions, $E_1$ can be conclusively distinguished from $E_2$ with nonzero probability.
}

Note, that the above theorem is symmetric in a sense that $E_2$ can also be conclusively distinguished from $E_1$ with nonzero probability, but there may be no common measurement that allows for simultaneous conclusive distinguishing $E_1$ from $E_2$ and $E_2$ from $E_1$ with nonzero probability.

In special case when at least one of the extremal boxes is local in case of 2 inputs and 2 outputs, again using lemma \ref{lem:ext-dis} we obtain the following fact:

{\theorem Any extremal bipartite box with two inputs of arbitrary cardinalities $d_A$ and $d_B$ and two outputs of arbitrary cardinalities $d'_A$ and $d'_B$ is perfectly distinguishable from any extremal local bipartite box of the same dimensions by CLP operation.}

{\it Proof}. 
Fix arbitrarily a pair: an extremal box $E$ and a local extremal box $L$.
Note, that in bipartite case of 2 inputs and 2 outputs any extremal local box is deterministic i.e. is a family of $d_A\times d_B$ distributions with single entry equal to 1, and all others zero. By lemma \ref{lem:ext-dis} for some measurement $x_0,y_0$, the support of distribution $P_L(a,b|x_0,y_0)$ is not contained within the support of $P_E(a,b|x_0,y_0)$ which means in this case, that these supports are disjoined. This implies that $L$ is conclusively distinguishable from $E$ and vice versa for the same measurement with probability 1, hence the probability of success of discrimination between them equals 1.\blacksquare

\section{Conclusions}
We have extended a paradigm of distinguishing entangled states to the world of boxes. We have considered distinguishing of isotropic boxes, and provided easy linear program that gives the bound on the probability of success of discrimination among them by means of completely locality preserving operations which discriminates the ensemble. As a corollary we obtained bounds for the probability of success of discrimination of maximally nonlocal boxes as well as isotropic boxes with the same parameter. The bound is obtained in terms non-local cost of special input box: the mixture of classically correlated copies of boxes that are to be discriminated. The key argument in this result was monotonicity of non-local cost under CLP operations. We have shown also an example of useful CLP operation which is the {\it comparing operation}: local measurement followed by communication of the results, grouping them according to some partition and tracing out the results. We proved that it can help in discriminating between pairs of extremal boxes in bipartite case for any pairs in $2\times 2$ case, or between any local extremal box and any other extremal local boxes in bipartite case of boxes with 2 inputs and 2 outputs of arbitrary cardinalities. 
It would be interesting if application of other monotone then non-locality cost would give better upper bounds. Note, that comparing operation is not the only
one possible for boxes, as e.g. one could apply wiring between the parties \cite{Allcocketal2009}. 

Finally, we have to stress, that presented upper bounds on
probability of success should be considered rather as a demonstration of analogy between entanglement and non-locality - two resource theories. This is because the bounds seems to be very rough, as most probably discriminating between two boxes perfectly by means of CLP e.g. between PR box and anti PR box, is the best strategy when one is given mixture of more than two maximally non-local boxes. This strategy yields probability of success equal to $2\over n$ when $8\geq n\geq 2$, which is far from obtained bounds. It would be then interesting if one could find more tight ones, perhaps using more direct approach by considering general form of LP operations \cite{Barrett-GPT} than via monotones presented here. 

\begin{acknowledgments}

We thank M. Horodecki, R. Horodecki and D. Cavalcanti for discussion and M.T. Quintino and P. Joshi for helpful comments.
This research is partially funded by QESSENCE grant and grants BMN nr 538-5300-0637-1 and 538-5300-0975-12.

\end{acknowledgments}

\appendix
\section{Proof of corollary \ref{cor:loc-pres}}
\label{app:proof}
We first need to check that control-$O_j$ operation preserves locality on special class of local boxes, that appear in our considerations. 
Namely, consider a local box 
\be
\sum_i p_i P_i(a,c|x,u)\ot P_i(b,d|y,v) 
\ee
where inputs $x$ and $y$ are unary. It is transformed by control-$O_j$ operation
into 
\be
\sum_i p_i P_i(a,h_a(c)|x, \tilde{h}_a(u))\ot P_i(b,g_b(d)|y,\tilde{g}_b(v)) 
\ee
where functions $h,\tilde{h},g,\tilde{g}$ are either identity or a bitflip respectively.
Hence, the output box is a mixture of local boxes, 
we only need to check that $P_i(a,h_a(c)|x, \tilde{h}_a(u))$ and $P_i(b,g_b(d)|y, \tilde{g}_b(v))$
are fully non-signaling. It holds, indeed, as unary input can not signal, while 
\be
\sum_c P_i(a,h_a(c)|x,\tilde{h}_a(u_0)) = \sum P_i(a,h_a(c)|x,\tilde{h}_a(u_1)) 
\ee
for any $a$ and values $u_0$ and $u_1$, as for fixed $a$ $\tilde{h}_a$ just permutes the inputs,
while $h_a$ changes order of summation, keeping the range of $c$, hence the thesis follows 
from non-signaling of box $P_i(ac|xu)$.

Next step is to show that control-$O_j$ operation transforms non-signaling boxes into non-signaling ones.
There are 5 inequivalent ways to distinguish a subsystem out of a box of the form 
\be
\sum_{ij} p(i,j) F^{(j)}(e|s)\ot F^{(j)}(f|t)\ot P_i(c,d |u,v)
\ee
where $P_i(c,d|u,v)$ are non-signaling boxes. After applying controlled-$O_j$ operation, there is:
\be
\sum_{ij} p(i,j) F^{(j)}(e|s)\ot F^{(j)}(f|t)\ot P_i(h_j(c),g_j(d) |\tilde{h}_j(u),\tilde{g}_j(v)).
\ee
We just show one example of full non-signaling condition, as the others follow similar lines.
Namely we show now that inputs $s$ and $u$ does not signal to systems $v$ and $t$. Indeed:
this condition reads
\begin{widetext}
\be
\forall_{s_0,u_0,u_1} \forall_{v_0,t_0,d_0,f_0}   
\sum_e\sum_c\sum_{ij}p(i,j)\delta_{j,e_0}\delta_{j,f_0} P_i(h_j(c),g_j(d)|\tilde{h}_j(u_0),\tilde{g}_j(v_0)) = LHS(v_1)
\ee

where $LHS(u_1)$ denotes the equation on RHS of equality with $u_1$ in place of $u_0$. This happens iff 

\be
\forall_{u_0,u_1} \forall_{v_0,d_0,f_0}   
\sum_c\sum_{i}p(i,f_0)P_i(h_{f_0}(c),g_{f_0}(d)|\tilde{h}_{f_0}(u_0),\tilde{g}_{f_0}(v_0)) = LHS(u_1)
\ee
\end{widetext}
But we observe, that for all $i$ and $j$ there is
\be
\sum_c P_i(h_{f_0}(c),g_{f_0}(d)|\tilde{h}_{f_0}(u_0),\tilde{g}_{f_0}(v_0)) = LHS(u_1)
\ee
which follows from non-signaling of the boxes $P_i(c,d|u,v)$ for each $i$, and that the functions
$h,\tilde{h},g,\tilde{g}$ are only bit-flips. 

Finally, we observe that control-$O_j$ operations is linear. It is easy to see, that partial trace of a subsystem,
and twirling operation are CLP operations. This ends the proof of corollary \ref{cor:loc-pres}.\blacksquare

\section{Comparing operations are CLP}\label{subsec:comparing-ops}

In this section we show that a {\it comparing operations} are valid CLP operations.
This operations transforms a box $P(ab|xy)$ into $\Lambda(P)$ given below, defined on systems $CDEF$, where
to fix the considerations we assume that measurement $x=i,y=j$ has been performed on initial box,
\be
\Lambda(P) := \sum_k \sum_{(a,b)\in I_k} P(ab|x=i,y=j) P^{(k)}_E(e|s)\otimes P^{(k)}_F(f|t)
\label{eq:exCLP}
\ee
The family $\{I_k\}_k$ is a partition of the set of all pairs of outputs $(a,b)$ into disjoined sets of pairs, specific to given comparing operation, and
$P^{(k)}_A(e|s)$ and $P^{(k)}_B(f|t)$ are the boxes with unary input $s=0,t=0$ and output probability distributions 
$\delta_{k,e}$ and $\delta_{k,f}$ respectively. In what follows we will write $x_i$,$y_j$,$s_0$,$t_0$ instead of $x=i$,$y=j$,$s=0$,$t=0$ respectively.
Note, that one can obtain $\Lambda(P)$ via exchanging results control operation, and tracing out the results.

\subsection{verifying CLP conditions}
We argue now, that operation (\ref{eq:exCLP}) is CLP. Note that it is enough to show, that $\Lambda \otimes I$ is LP, as from it, we have immediately that
$\Lambda$ itself is LP. Indeed, suppose it is not the case, that is $\Lambda_A$ is not LP on some box $P(a|x)$. We have then $\Lambda_A(P(a|x)) = \Lambda_A\ot I_B(P(a|x)\otimes P(b|y))$ where $P(b|y)$ is a trivial box on system B: with 1 input, and 1 output, with probability 1, because $P(a|x)P(b|y) = P(a|x)$ in this case. This however implies that $\Lambda_A$ is not CLP, which is desired contradiction.

Consider then a box
\be
M = P(a,b,\bar{c},\bar{d}|x,y,\bar{u},\bar{v})
\ee
on systems $ABCD$ with $C= C_1,...,C_n$ and $D=D_1,...,D_n$.
We now apply $\Lambda_{AB}\ot I_{CD}$. Resulting box is on systems $CDEF$:
\begin{multline}
\Lambda\ot I (M) = \nonumber \\
\sum_k \sum_{(a,b)\in I_k} P(a,b,\bar{c},\bar{d}|x_i,y_j,\bar{u},\bar{v})\otimes P^{(k)}_E(e|s)\otimes P^{(k)}_F(f|t)
\end{multline}

We have to prove now the list of features (i)-(iv) given in definition \ref{def:LP}.
To prove validity of $\Lambda \otimes I$ it is enough to notice that fixing $\bar{u}_0,\bar{v}_0$ $s_0$ and $t_0$ and summing over outputs we get 
\be
\sum_{\bar{c},\bar{d},e,f} \sum_k \sum_{(a,b)\in I_k} P(a,b,\bar{c},\bar{d}|x_i,y_j,\bar{u}_0,\bar{v}_0) P^{(k)}_E(e|s_0)P^{(k)}_F(f|t_0)
\ee
that equals
\be
\sum_{\bar{c},\bar{d},e,f} \sum_k \sum_{(a,b)\in I_k} P(a,b,\bar{c},\bar{d}|x_i,y_j,\bar{u}_0,\bar{v}_0)
\ee
which is desired 1, since initial box was valid for input $x_i,y_j,\bar{u}_0,\bar{v}_0,s_0,t_0$.

To prove linearity, we observe that if we allow mixture of boxes $M =\sum_{l} \alpha_l P_l(a,b,\bar{c},\bar{d}|x,y,\bar{u},\bar{v})$, then 
the result $\Lambda\ot I(M)$ would be
\be
\sum_k \sum_{(a,b)\in I_k} [\sum_{l}\alpha_l P_l(a,b,\bar{c},\bar{d}|x_i,y_j,\bar{u},\bar{v})] P^{(k)}_E(e|s)\otimes P^{(k)}_F(f|t)
\ee
which is the same as 
\begin{multline}
\sum_{l}\alpha_l \Lambda\ot I(M_l) = \nonumber \\
\sum_{l}\alpha_l [\sum_k \sum_{(a,b)\in I_k} P(a,b,\bar{c},\bar{d}|x_i,y_j,\bar{u},\bar{v}) P^{(k)}_E(e|s)\otimes P^{(k)}_F(f|t)]
\end{multline}
since we can change the order of summation.

The argument that operation $\Lambda \ot I$ preserves non-signaling is more demanding.
To show the full non-signaling we need to prove two conditions:
\ben
C^ID^J\not\hspace{-1.3mm}{\leftrightarrow}EC^{N-I}FD^{N-J} \label{eq:first-ns}\\
EC^ID^J\not\hspace{-1.3mm}{\leftrightarrow}C^{N-I}FD^{N-J} \label{eq:sec-ns}
\een
where $I,J\subset \{1,...,n\}\equiv N$ and we do not consider the case when $I$ and $J$ are empty at the same time. (Note that the cases 
\ben 
C^IFD^J\not\hspace{-1.3mm}{\leftrightarrow}EC^{N-I}D^{N-J}, \nonumber \\
EC^IFD^J \not\hspace{-1.3mm}{\leftrightarrow}C^{N-I}D^{N-J} \nonumber
\een
are covered by the first two above).
We will show (\ref{eq:first-ns}) only, as (\ref{eq:sec-ns}) follows from analogous considerations. 
In what follows, for any multivariable named $\bar{w}\equiv (w^1,...,w^n)$, by $w^I$ we mean the variables with indices indicated by set of indices $I\subseteq N=\{1,...,n\}$. By $\bar{w}_0^I$ we mean the variables $w^i$ fixed to some values $w^i_0$ each for all $i\in I$ and 
by $\bar{w}_0{\bs}w^I$ we mean that for all $i \notin I$ variables  $w^i$ are fixed to some values $w^i_0$, but for $i\in I$ they are not fixed. 
Note, that in what follows we never put $s$ and $t$ under universal quantifier, since they have single value, we only fix them to $s_0$ and $t_0$ properly. 
To satisfy the non-signaling condition which we now focus on, there should be:
\begin{widetext}
\begin{multline}
\forall_{e_0,f_0,\bar{c}_0{\bs}c^I,\bar{d}_0{\bs}d^I,\bar{u}_0{\bs}u^I,\bar{v}_0{\bs}v^J}\quad \forall_{\bar{u}_0^{I},\bar{u}_1^{I},\bar{v}_1^{J},\bar{v}_1^{J}}   \\
\sum_{c^I,d^J} \sum_k \sum_{(a,b)\in I_k} P(a,b,\bar{c}_0{\bs}c^I,c^I,\bar{d}_0{\bs}d^J,d^J|x_i,y_j,\bar{u}_0{\bs}u^I,\bar{u}_0^I,\bar{v}_0{\bs}v^J,\bar{v}_0^J) P^{(k)}_E(e_0|s_0)P^{(k)}_F(f_0|t_0)  = \text{LHS}(\bar{u}_1^{I},\bar{v}_1^{J})
\end{multline}

where $LHS(\bar{u}_1^{I},\bar{v}_1^{J})$ denotes left-hand-side of the equation with $\bar{u}_1^{I}$ in place of $\bar{u}_0^{I}$ and $\bar{v}_1^{I}$ in place of $\bar{v}_0^{I}$.
Due to definition of $P^{(k)}_F(f|t_1) $ and $P^{(k)}_E(e|s_1) $ we have that LHS of the above equation equals 0 if $e_0\neq f_0$, and so equals RHS then,
while for $e_0=f_0$ the above set of equations reduces to:
\begin{multline}
\forall_{e_0,\bar{c}_0{\bs}c^I,\bar{d}_0{\bs}d^I,\bar{u}_0{\bs}u^I,\bar{v}_0{\bs}v^J}\quad \forall_{\bar{u}_0^{I},\bar{u}_1^{I},\bar{v}_1^{J},\bar{v}_1^{J}}  \\
\sum_{c^I,d^J} \sum_{(a,b)\in I_{e_0}} P(a,b,\bar{c}_0{\bs}c^I,c^I,\bar{d}_0{\bs}d^J,d^J|x_i,y_j,\bar{u}_0{\bs}u^I,\bar{u}_0^I,\bar{v}_0{\bs}v^J,\bar{v}_0^J)=LHS(\bar{u}_1^{I},\bar{v}_1^{J})
\end{multline}
which happens for all choice of variables that we can vary over, since for any fixed $(a_0,b_0) \in I_{e_0}$ there is 
\be
\sum_{c^I,d^J} P(a_0,b_0,\bar{c}_0{\bs}c^I,c^I,\bar{d}_0{\bs}d^J,d^J|x_i,y_j,\bar{u}_0{\bs}u^I,\bar{u}_0^I,\bar{v}_0{\bs}v^J,\bar{v}_0^J) P^{(k)}_E(e_0|s_0)P^{(k)}_F(f_0|t_0) = \text{LHS}(\bar{u}_1^{I},\bar{v}_1^{J})
\ee
due to non-signaling $C^ID^J\not\hspace{-1.3mm}{\rightarrow}AC^{N-I}BD^{N-J}$ of the original box $M$.
To prove the converse non-signaling condition we need to show the following equalities:
\begin{multline}
\forall_{\bar{c}_0^{I},\bar{d}_0^{J},\bar{u}_0^{I},\bar{v}_0^{J}} \quad
\forall_{\bar{u}_0{\bs}u^{I},\bar{u}_1{\bs}u^{I},\bar{v}_0{\bs}v^{J},\bar{v}_1{\bs}v^{J}} \\
\sum_{e,f,\bar{c}{\bs}c^I,\bar{d}{\bs}d^I} \sum_k \sum_{(a,b)\in I_k} P(a,b,\bar{c}{\bs}c^I,c_0^I,\bar{d}{\bs}d^J,d_0^J|x_i,y_j,\bar{u}_0{\bs}u^I,u_0^I,\bar{v}_0{\bs}v^J,\bar{v}_0^J) P^{(k)}_E(e|s_0)P^{(k)}_F(f|t_0)  = \text{LHS}(\bar{u}_1^{I}{\bs}u^{I},\bar{v}_1^{J}{\bs}v^J)
\end{multline}
again, we notice, that we need to prove 
\begin{multline}
\forall_{\bar{c}_0^{I},\bar{d}_0^{J},\bar{u}_0^{I},\bar{v}_0^{J}} \quad
\forall_{\bar{u}_0{\bs}u^{I},\bar{u}_1{\bs}u^{I},\bar{v}_0{\bs}v^{J},\bar{v}_1{\bs}v^{J}} \\
\sum_{\bar{c}{\bs}c^I,\bar{d}{\bs}d^I} \sum_k \sum_{(a,b)\in I_k} P(a,b,\bar{c}{\bs}c^I,c_0^I,\bar{d}{\bs}d^J,d_0^J|x_i,y_j,\bar{u}_0{\bs}u^I,u_0^I,\bar{v}_0{\bs}v^J,\bar{v}_0^J)  = \text{LHS}(\bar{u}_1^{I}{\bs}u^{I},\bar{v}_1^{J}{\bs}v^J)
\end{multline}
which is true, as it follows from non-signaling condition $AC^{N-I}BD^{N-J}\not\hspace{-1.3mm}{\rightarrow}C^ID^J$ of the original box $M$. Thus
we have proved (\ref{eq:first-ns}). 

Finally we need to prove that $\Lambda\ot I$ preserves locality. 
To this end consider a local box
\be
\sum_{\lambda} p(\lambda) P^{\lambda}(a\bar{c}|x\bar{u})\otimes P^{\lambda}(b\bar{d}|y\bar{v})
\ee
It is transformed into
\be
\sum_{\lambda} \sum_k \sum_{(a,b)\in I_k^{\lambda}} p(\lambda) [P^{(k)}_E(e|s)\otimes P^{\lambda}(a\bar{c}|x\bar{u})]\otimes [P^{(k)}_F(f|t)\otimes P^{\lambda}(b\bar{d}|y\bar{v})]
\label{eq:last-form}
\ee
by definition of locality, there are well defined normalization factors:
\ben
N_{EC}^{(\lambda,a)} = \sum_{\bar{c}} P^{\lambda}(a,\bar{c}|x=i,\bar{u}_0) \\
N_{FD}^{(\lambda,b)} = \sum_{\bar{d}} P^{\lambda}(b,\bar{d}|y=j,\bar{v}_0)
\een
so that our box in (\ref{eq:last-form}) looks like

\be
\sum_{\lambda,k,(a,b)\in I^{\lambda}_k} p(\lambda)N_{EC}^{(\lambda,a)}N_{FD}^{(\lambda,b)}
[P^{(k)}_E(e|s)\otimes {1\over N_{EC}^{(\lambda,a)}} P^{\lambda}(a\bar{c}|x\bar{u})]\otimes [P^{(k)}_F(f|t)\otimes {1\over N_{FD}^{(\lambda,b)}}P^{\lambda}(b\bar{d}|y\bar{v})]
\label{eq:box}
\ee

to see that this is a valid $LR_{ns}$ box, consider a random variable $\lambda'$ with a distribution defined for all $(\lambda,k,a,b)$ as 

\be
\left\{\begin{array}{l} 
\lambda'(\lambda,k,a,b) = p(\lambda)N_{EC}^{(\lambda,a)}N_{FD}^{(\lambda,b)} \quad \text{for} \quad (a,b)\in I^{\lambda}_k \\
\lambda'(\lambda,k,a,b) = 0 \quad \text{else}
\end{array}\right.
\ee

Note that this is well defined distribution of a random variable over Cartesian product of ranges of $\lambda$, $k$ and ranges of $a$ and $b$.
Indeed, 
\be
\sum_{\lambda,k,a,b} \lambda'(\lambda,k,a,b)= \sum_{\lambda} \sum_{k}\sum_{(a,b)\in I^{\lambda}_k} p(\lambda)N_{EC}^{(\lambda,a)}N_{FD}^{(\lambda,b)}
\ee
which is nothing but
\be
\sum_{\lambda,a,b} p(\lambda) \sum_{\bar{c},\bar{d}}P^{\lambda}(a\bar{c}|x_i,\bar{u}_0)P^{\lambda}(b\bar{d}|y_j,\bar{v}_0),
\ee
and equals 1, since it is the distribution of outcomes of  measurement $x_i,y_j,\bar{u}_0\bar{v}_0$ on the original box $M$.
Now we can rewrite the box (\ref{eq:box})
\be
\sum_{\lambda,k,a,b}  \lambda'(\lambda,k,a,b) 
[X_{EC}^{(\lambda,k,a,b)}]\otimes [Y_{FD}^{(\lambda,k,a,b)}]
\ee
where $X_{EC}^{(\lambda,k,a,b)}=P^{(k)}_E(e|s)\otimes {1\over N_{EC}^{(\lambda,a)}} P^{\lambda}(a\bar{c}|x\bar{u})$ 
and  $Y_{FD}^{(\lambda,k,a,b)}= P^{(k)}_F(f|t)\otimes {1\over N_{FD}^{(\lambda,b)}}P^{\lambda}(b\bar{d}|y\bar{v})$ are legitimate boxes
on Alice's and Bob's system respectively. It is also easy to see that the boxes $[X_{EC}^{(\lambda,k,a,b)}]$ and $[Y_{FD}^{(\lambda,k,a,b)}]$
are fully non-signaling, as the original box was $LR_{ns}$. Hence we proved that the output of $\Lambda\ot I$ acting on $LR_{ns}$ box
is an $LR_{ns}$ box.

\end{widetext}

\bibliographystyle{apsrev}

%\bibliography{rmp11-hugekey-phd}

\end{document}